\documentclass[11pt,a4paper]{elsarticle}
\makeatletter
\def\ps@pprintTitle{%
 \let\@oddhead\@empty
 \let\@evenhead\@empty
 \def\@oddfoot{\centerline{\thepage}}%
 \let\@evenfoot\@oddfoot}
\makeatother

\usepackage{amsfonts}
\usepackage{amsmath}
\usepackage{geometry}
\usepackage{graphicx}
\usepackage{amssymb}
\usepackage{amsmath}
\usepackage{amsfonts}
\usepackage{amssymb}
\usepackage{color}
\usepackage{psfrag}

\newtheorem{theorem}{Theorem}

\newtheorem{proposition}[theorem]{Proposition}

\begin{document}

\begin{frontmatter}

\title{Computation of the Madelung constant for hypercubic crystal structures in any dimension}
\author{Malik Mamode}
\address{Department of Physics - Laboratoire PIMENT - University of La R\'{e}union (France)}
\ead{malik.mamode@univ-reunion.fr}

\begin{abstract}
A new method of computing the Madelung constants for hypercubic crystal structures in any dimension $n\geq 2$ is given. It is shown for $n\geq 3$ that the Madelung constant may be obtained in a simple, efficient and unambiguous way as the Hadamard finite part of the integral representation of the potential within the crystal which is divergent at any point charge location. Such a regularization method fails in the bidimensional case due to the logarithmic nature of singularities for the potential. In that case, a specific approach is proposed taking in account the scale invariance of the Poisson equation and the existence of a finite horizon for each point charge in the plane. Since a closed-form exact solution for the 2D electrostatic potential may be derived, one shows that the Madelung constant may be defined via an appropriate limit calculation as the mean value of potential energies of charges composing the unit cell.
\end{abstract}


\end{frontmatter}

\section*{Introduction and Notation}

The Madelung constant of ionic solids pertains to the geometry of the crystal structure and electrostatic interactions between the charges. This is a key feature determining the electrostatic energy in lattices and involved in the study of their stability. As an example, if we consider point charges $(-1)^{k+m+p}$ arranged over the lattice $\mathbb{Z}^{3}$ (it is the case for the NaCl crystal using the common convention of crystallography that the nearest-neighbour separation is unity, a cation Na$^{+}$ being placed at the origin), the related Madelung constant is given by the triple sum of $3$-dimensional Coulomb terms over all integers except the term $k=m=p=0$,
\begin{equation}\label{MadelungNaCl}
M_{\mathrm{NaCl}}=\sum_{\vert k\vert +\vert m\vert +\vert p\vert \neq 0}\dfrac{(-1)^{k+m+p}}{\left( k^{2}+m^{2}+p^{2}\right)^{1/2} } .
\end{equation}
In Gaussian units, the latter constant represents the potential seen by the origin charge, its self-potential being removed and hence is equal to the potential energy in which this reference charge resides. Obviously here, the Madelung constant has the same value at any other lattice site so that the total electrostatic energy per unit cell of the NaCl crystal (which contains four positive charges and four negative charges) is
\[
U=\dfrac{1}{2}\times 8M_{\mathrm{NaCl}}=4M_{\mathrm{NaCl}}.
\]
More generally, the usual definition of Madelung constants for crystal lattices involves summing the contributions at the origin of all charges. This leads to lattice sums similar to (\ref{MadelungNaCl}) whose numerical evaluation for physical applications is almost always problematic since such infinite series are conditionnally convergent and have not the arrangement property. The study of convergence of such series is yet to date a mathematically challenging problem. Several approaches have been proposed for giving a precise and unambiguous meaning for these lattice sums and hence for computing the Madelung constants with higher accuracy (see e.g. \cite{borwein_book,Linton,Crandall2} and references therein for details). In particular, (\ref{MadelungNaCl}) is an example of Epstein zeta function whose a fast evaluation is proposed in \cite{Crandall3} (see also \cite{Tyagi}) leading to the numerical value
\begin{equation}\label{Madelung_NaCl}
M_{\mathrm{NaCl}}=-1.747 \ 564 \ 594 \ 633 \ 182 \ 190 \ 636 \ 212 \ 035 \ldots 
\end{equation}
which is exact at more than sixty decimal digits. In contrast, a closed-form expression for $M_{\mathrm{NaCl}}$ (and the Madelung constants in general) remains to date untraceable.

The present paper is concerned with the Madelung constant for invariant $n$-dimensional cubic crystals ($n\geq 2$) structured over a Bravais lattice $\Lambda =2a\mathbb{Z}\times \ldots \times 2a\mathbb{Z}$ ($a>0$). For this, we consider the $n$-dimensional flat torus $\mathbb{T}^{n}=\mathbb{R}^{n}/\Lambda$, quotient of the Euclidean space $\mathbb{R}^{n}$ by the lattice $\Lambda$ which provides thereby the appropriate geometrical framework to tackle the electrostatics of the crystal, in particular when solving the Poisson equation $\Delta V=-4\pi \rho$ for any periodic point charge distribution $\rho$. In this setting, a fundamental domain of the torus corresponds to a Bravais cell (hypercubic unit cell) of measure $\vert T\vert =2^{n}a^{n}$.

In Section 1, we determine the fundamental solution $\Psi$ for the Laplacian on the $n$-torus i.e. the unique distribution (up to a constant) solution of 
\begin{equation}\label{G2}
\Delta \Psi = \delta (\mathbf{x})-\dfrac{1}{\vert T\vert} \ \ \ \mathrm{on} \ \ \mathbb{T}^{n}
\end{equation}
where $\delta$ is the Dirac distribution at the origin. As an application, we may deduce in Section 2 given the charge distribution $\rho$ in the unit cell, the analytical exact expression for the potential within the crystal lattice as the multiple convolution product on the torus,
\begin{equation}\label{AA}
V=-4\pi \Psi \overset{\textbf{x}}{\ast}\rho .
\end{equation}
Such a result allows thus to revisit in the two last sections the definition of the Madelung constant for any invariant hypercubic lattice complexes in any dimension $n\geq 2$. In contrast with the usual approach by conditionally convergent lattice sums, we show in Section 3 that the Madelung constant can be defined unambiguously when $n\geq 3$ as the Hadamard finite part of a divergent integral involving Jacobi theta functions whose evaluation remains easy and sufficiently accurate for a practical use. Two application examples (for the NaCl and CsCl crystal) show that such a definition gives results which are consistent with traditional ones.

In Section 4, we discuss for ending the very singular case $n=2$ (planar square crystal) and the impossibility to define similarly the Madelung constant owing to the existence of logarithmic singularities for the potential at each point charge location. In the $2$-dimensional case, the regularization proposed above is not unique and arbitrary. We shall see that this difficulty is closely related to the scale invariance of the fundamental solution for the Laplacian in the whole plane. The consequences for electrostatics are thus ($i$) the impossibility to fix an infinite horizon for the $2$-Coulomb potential due to an isolated charge in the plane $\mathbb{R}^{2}$ (we call horizon the distance which separates the point charge to a reference point arbitrarily chosen where the potential is zero, see \cite{Mam}), ($ii$) the impossibility to define in a satisfactory way the Madelung constant of a planar square crystal by a lattice sum like (\ref{MadelungNaCl}) or relatives. Nevertheless, since there exists a closed-form solution for the potential within the crystal, we show considering the NaCl and CsCl crystals that the related constant may be defined as the mean electrostatic potential energies of charges per unit cell and that it is possible to compute it in an easy and exact way.

\section{Fundamental solution}

We set the $n$-dimensional theta function
\begin{equation}\label{theta}
\Theta (\mathbf{x}\vert v)=\vartheta_{3}\left( \dfrac{\pi x_{1}}{2a}\right. \left| i\dfrac{\pi v}{a^{2}}\right)\ldots \vartheta_{3}\left( \dfrac{\pi x_{n}}{2a}\right. \left| i\dfrac{\pi v}{a^{2}}\right) \ \ , \ \mathbf{x}=\left( x_{1},\ldots x_{n}\right) \in \mathbb{R}^{n} \ \ , \ v>0
\end{equation}
where $\vartheta_{3}$ denotes the third Jacobi theta function which is the exponentially convergent series \cite{Grad}
\begin{equation}\label{SerieFourier}
\vartheta_{3}\left( \dfrac{\pi x_{l}}{2a}\right. \left| i\dfrac{\pi v}{a^{2}}\right)=\sum_{k_{l}=-\infty}^{+\infty}e^{-\pi^{2}k_{l}^{2}v/a^{2}}e^{i \pi k_{l} x_{l}/a}.
\end{equation}
It ensues that (\ref{theta}) may be given by the following lattice sum 
\begin{equation}\label{SerieFourier1}
\Theta (\mathbf{x}\vert v)=\sum_{\mathbf{k}\in\mathbb{Z}^{n} }e^{-\pi^{2}\Vert \mathbf{k}\Vert^{2}v/a^{2}}e^{i\pi\mathbf{x}.\mathbf{k}}.
\end{equation}
Recall that $\Theta$ is solution of the heat equation for $\mathbf{x}\in \mathbb{R}^{n}$ and the modular parameter $v>0$ \cite{Grad}, 
\begin{equation}\label{heat1}
\left( \Delta -\dfrac{\partial}{\partial v}\right) \Theta (\mathbf{x}\vert v)=0 \ \ \ , \ \ \Delta =\sum_{l=1}^{n}\dfrac{\partial^{2}}{\partial x_{l}^{2}}.
\end{equation}
In addition, for any fixed $v>0$, $\Theta$ defines a (multiply-periodic) distribution on the $n$-torus $\mathbb{T}^{n}$ such that in a distributional sense,
\begin{equation}\label{lim1}
 \underset{v \rightarrow +\infty}{\mathrm{lim}}\Theta =1 \ \ \ \mathrm{and} \ \ \underset{v \rightarrow 0+}{\mathrm{lim}}\Theta=\vert T\vert \delta(\mathbf{x}),
\end{equation}
$\delta$ being the Dirac distribution on $\mathbb{T}^{n}$. In particular, at any lattice point, the function $\Theta$ is $O(v^{-n/2})$ as $v$ tends to zero (see below, (\ref{asympt})).

Hence, we can prove that
\begin{proposition}
The distribution defined by the integral
\begin{equation} \label{FS}
\Psi (\mathbf{x}):=\dfrac{1}{\vert T\vert}\int_{0}^{+\infty} \left\lbrace 1-\Theta(\mathbf{x}\vert v)\right\rbrace  dv 
\end{equation}
is the fundamental solution for the Laplacian on the $n$-torus such that
\begin{equation}\label{compati2}
\int \ldots\int_{\mathbb{T}^{n}}\Psi(\mathbf{x})d\mathbf{x}=0.
\end{equation}
\end{proposition}
Indeed, notice first that $\Psi$ has a singularity at each point of the lattice $\Lambda$ for $n\geq 2$ (logarithmic singularity if $n=2$) by virtue of the above remark. On the other hand for $\mathbf{x}\notin\Lambda$ and by splitting the integral (\ref{FS}) into $\int_{0}^{1}+\int_{1}^{+\infty}$, one can deduce that the integral $\int_{0}^{1}$ is convergent since (see (\ref{lim1})),
\[
\Theta(\mathbf{x}\vert v)\ \rightarrow 0  \ \ \ \mathrm{as} \ \ \ v\rightarrow 0+ .
\]
Then, using (\ref{SerieFourier1}), one has for any $\mathbf{x}$,
\begin{eqnarray*}
\left| 1-\Theta(\mathbf{x}\vert v) \right| &\leq & \sum_{\mathbf{k}\in \mathbb{Z}^{n}\setminus \mathbf{\lbrace 0\rbrace}}e^{-\pi^{2}\mathbf{k}^{2}v/a^{2}}=2n \ e^{-\pi^{2}v/a^{2}}+\sum_{\mathbf{k}^{2}\geq 2}e^{-\pi^{2}\mathbf{k}^{2}v/a^{2}}
\\ 
&\leq & 2n \ e^{-\pi^{2}v/a^{2}}+\int_{1}^{+\infty}S_{n-1}(\rho)\ e^{-\pi^{2}\rho^{2}v/a^{2}} d\rho
\end{eqnarray*}
where $S_{n-1}(\rho)=2\pi^{n/2}\rho^{n-1}/\Gamma(n/2)$ is the surface area of the $(n-1)$-sphere in $\mathbb{R}^{n}$ of radius $\rho$. Hence,
\begin{equation}\label{maj1}
\left| 1-\Theta(\mathbf{x}\vert v) \right|\leq 2n \ e^{-\pi^{2}v/c^{2}}+\dfrac{\pi^{n/2}}{\Gamma\left( \dfrac{n}{2}\right) }E_{1-\frac{n}{2}}\left( \dfrac{\pi^{2}v}{c^{2}}\right) 
\end{equation}
where $E_{\nu}(z)$ denotes the Exponential Integral such that \citep{Abram}
\begin{equation}\label{EI}
E_{\nu}(z)=\dfrac{e^{-z}}{z}\left( 1+O\left( \dfrac{1}{z}\right) \right) \ \ \ \mathrm{as} \ \ \vert z \vert \rightarrow +\infty .  
\end{equation}
Consequently, the integral $\int_{1}^{+\infty}$ is absolutely convergent, hence also the integral (\ref{FS}) for any $\mathbf{x}\notin\Lambda$.

To end the proof, it remains to show that $\Psi$ is solution of (\ref{G2}) on the $n$-torus. For this, it suffices to notice that
\begin{equation*}
\Delta \Psi(\mathbf{x},\lambda)=-\dfrac{1}{\vert T\vert}\int_{0}^{+\infty} \Delta\Theta(\mathbf{x}\vert v)  dv
=-\dfrac{1}{\vert T\vert}\int_{0}^{+\infty} \dfrac{\partial}{\partial v}\Theta(\mathbf{x}\vert v)  dv
\end{equation*}
and finish the calculation using (\ref{lim1}). The formula (\ref{compati2}) results from the property that the $\vartheta_{3}$-functions are of zero mean value and it will be the same for $\Psi$. $\blacksquare$

It is worth to note that the integral (\ref{FS}) may be related to a multidimensional zeta function. Indeed, consider the Mellin transform,
\begin{equation}\label{zeta}
\mathcal{E}(\mathbf{x},s):=\int_{0}^{+\infty}v^{-1+s/2} \left\lbrace 1-\Theta(\mathbf{x}\vert v)\right\rbrace  dv .
\end{equation}
All the above results justify that the latter integral is absolutely convergent - and the so defined function $\mathcal{E}$ is analytic in the complex variable $s$ - when $\Re s>0$ given any fixed $\mathbf{x}\notin \Lambda$. Using (\ref{SerieFourier1}) and by interchanging integration and summation, we easily show that (\ref{zeta}) is formally  the $n$-dimensional Epstein zeta function expressed by the series,
\begin{equation}\label{zetaSF}
\mathcal{E}(\mathbf{x},s)=-\dfrac{\Gamma \left( \dfrac{s}{2}\right) }{\pi^{s}}\sum_{\mathbf{k}\in \mathbb{Z}^{n}\setminus \mathbf{\lbrace 0\rbrace}}\dfrac{e^{i\pi\mathbf{x}.\mathbf{k}}}{\Vert\mathbf{k}\Vert^{s}} .
\end{equation}
As it is known (see e.g. \cite{Linton,Crandall2}), such a series converges absolutely for $\Re s>n$. In the contrary case, it has not the rearrangement property but its limit (here given by the integral (\ref{zeta}) which corresponds thus to its analytic continuation over the half-plane $\Re s>0$) is unique and must result from an appropriate ordering of the terms (see applications below). In particular, taking $s=2$, the fundamental solution for the Laplacian on the $n$-torus is the following zeta function given by the conditionally convergent lattice sum (Fourier series) for $n\geq 2$,
\begin{equation} \label{psi}
\Psi (\mathbf{x})=\dfrac{1}{\vert T\vert}\mathcal{E}(\mathbf{x},2)=-\dfrac{1 }{\pi^{2}\vert T\vert}\sum_{\mathbf{k}\in \mathbb{Z}^{n}\setminus \mathbf{\lbrace 0\rbrace}}\dfrac{e^{i\pi\mathbf{x}.\mathbf{k}}}{\Vert\mathbf{k}\Vert^{2}}.
\end{equation}

Let us consider two application examples.
\subsection{The $1$-torus ($n=1$)}

On one hand, $\Psi (x_{1})$ is the uniformly convergent series \cite{Grad}
\begin{equation}\label{psi0}
-\dfrac{a}{\pi^{2}}\sum_{k=1}^{+\infty}\dfrac{\cos\left( k\pi x_{1}/a\right) }{k^{2}}=-\dfrac{a}{6}+\dfrac{x_{1}}{2}-\dfrac{x_{1}^{2}}{4a} \ \ \ \mathrm{if} \ \ 0\leq x_{1}\leq 2a
\end{equation}
solution (in the sense of distributions) of the one-dimensional equation $\Psi''(x_{1})=\delta (x_{1})-1/2a$ on $\mathbb{T}^{1}$ i.e. on the circle of length $2a$. On the other hand, from (\ref{FS}),  this fundamental solution may be also related to an Hurwitz zeta function as \cite{Grad}
\begin{eqnarray*}
\Psi (x_{1})&=&\dfrac{1}{2a}\int_{0}^{+\infty}\left\lbrace 1-\vartheta_{3}\left( \dfrac{\pi x_{1}}{2a}\right. \left| i\dfrac{\pi v}{a^{2}}\right)\right\rbrace dv
\\
&=&a_{1}\left\lbrace \zeta \left( -1,\dfrac{x_{1}}{2a}\right) +\zeta \left( -1,1-\dfrac{x_{1}}{2a}\right)\right\rbrace 
\end{eqnarray*}
which can be expressed more simply in term of the second-order Bernoulli polynomial as
\[
\Psi (x_{1})=-\dfrac{a}{2}\left\lbrace B_{2}\left( \dfrac{x_{1}}{2a}\right) +B_{2}\left( 1-\dfrac{x_{1}}{2a}\right)\right\rbrace .
\]
All these results are valid for $0\leq x_{1}\leq 2a$.

\subsection{The $2$-torus ($n=2$)}

The series (\ref{psi}) is for instance arranged as
\[
\sum_{k_{1}\geq 1, k_{2}=0}+ \sum_{k_{2}\geq 1, k_{1}=0}+\sum_{k_{1}\geq 1}\left( \sum_{k_{2}\geq 1}\right) .
\]
Computing the repeated series by the standard summation formula \cite{Grad}
\[
\sum_{k=1}^{+\infty}\dfrac{\cos (kx)}{k^{2}+\alpha^{2}}=\dfrac{\pi}{2\alpha}\dfrac{\cosh(\alpha(\pi -x))}{\sinh(\alpha \pi)}-\dfrac{1}{2\alpha^{2}} \ \ \ \ \ \mathrm{for} \ \ 0\leq x\leq 2\pi
\]
one obtain straightforwardly the Fourier series
\begin{equation}\label{solfond3}
\Psi (x_{1},x_{2})=-\dfrac{1}{2\pi}\sum_{k_{1}=1}^{+\infty}\dfrac{\cosh(\pi k_{1}(a-x_{2})/a)\cos (\pi k_{1}x_{1}/a)}{k_{1}\sinh (\pi k_{1})}-\dfrac{1}{12}+\dfrac{x_{2}}{4a}-\dfrac{x_{2}^{2}}{8a^{2}}
\end{equation}
(or a similar expression by interchanging the roles of variables $x_{l}$) valid and uniformly convergent in any compact subset of the fundamental square domain $(0,2a)\times (0,2a)$ of the flat torus excluding the four vertices.

Now, let us remark that the latter series may be related to the series expansion of the logarithmic derivative of a theta function whence we may write \textit{up to a constant} \cite{Abram},
\[
-\dfrac{1}{2\pi}\sum_{k_{1}=1}^{+\infty}\dfrac{\cosh(\pi k_{1}x_{2}/a)\cos (\pi k_{1}x_{1}/a)}{k_{1}\sinh (\pi k_{1})}\doteq\dfrac{1}{2\pi} \log \left|  \vartheta_{4}\left( \left. \frac{\pi}{2a}(x_{1}+ix_{2}) \right| i \right)\right|
\]
where the symbol $\doteq$ means equality up to an additive constant and $\vartheta_{4}$ the fourth Jacobi theta function. Then, changing $x_{2}$ into $(x_{2}-a)$ and using a standard translation formula, we readily obtain the analytical expression of the fundamental solution for the Laplacian on the $2$-torus as,
\begin{equation} \label{psi1}
\Psi(x_{1},x_{2})=\dfrac{1}{4a^{2}}\mathcal{E}(x_{1},x_{2},2)\doteq\frac{1}{2\pi}\log \left|  \vartheta_{1}\left( \left. \frac{\pi}{2a}(x_{1}+ix_{2}) \right| i \right)\right| -\dfrac{x_{2}^{2}}{8a^{2}}
\end{equation}
$\vartheta_{1}$ denoting the first Jacobi theta function, or an identical expression by interchanging the roles of $x_{l}$.
This is exactly the result previously obtained by the author in \cite{Mam} wherein the Eisenstein's approach to elliptic functions via infinite series over lattices in the complex plane had been considered \cite{Walker}.

To our knowledge, for higher dimension $n\geq 3$, there exists to date no compact closed-form expression for the fundamental solution $\Psi$ apart from the analytical integral representation (\ref{FS}) which nonetheless suffices to bring a new light on the physical problem we are interested. 

\section{Potential within the crystal}

We detail in this section the analytical calculation of the potential within an invariant hypercubic lattice crystal characterized by a known distribution of point charges. For sake of simplicity, we restrict ourselves to two types of crystal structure: the NaCl-type crystal and the CsCl-type crystal in any dimension, the calculation being easily replicable for any other hypercubic structures. The Gaussian units are used.

\subsection{NaCl crystal}

Consider a $n$-dimensional NaCl-type crystal characterized by the following charge distribution on the flat torus $\mathbb{T}^{n}$,
\[
\rho (\textbf{x})=\sum_{\textbf{k}\in \lbrace0,1\rbrace^{n}}(-1)^{\vert\textbf{k}\vert}\delta (\textbf{x}-a\textbf{k}),
\]
where $\vert\textbf{k}\vert \in \lbrace 0,1,\ldots ,n\rbrace$ denotes the sum of components of the multi-index $\textbf{k}$. The unit cell defined on a fundamental domain of the torus contains thus $2^{n-1}$ point charge $+1$ for even $\vert\textbf{k}\vert$ and $2^{n-1}$ point charge $-1$ for odd $\vert\textbf{k}\vert$ such that the total charge in the cell is zero. The electrostatic potential $V$ within the crystal is thereby the multiply periodic distribution satisfying the Poisson equation $\Delta V=-4\pi \rho$ on the $n$-torus and whose existence is well guaranteed by the neutrality condition $\int \ldots\int_{\mathbb{T}^{n}}\rho \ d\textbf{x}=0$. As seen above, the potential is known up to a constant unless additional conditions are prescribed.

Indeed, considering (\ref{G2}) which defines the fundamental solution and the Poisson equation, one obtains
\[
V(\textbf{x})=-4\pi \ \Psi \overset{\textbf{x}}{\ast}\rho + V_{m}
\]
where $V_{m}$ is an arbitrary constant equal to the mean value of the potential viz.
\[
V_{m}=\dfrac{1}{\vert T\vert}\overset{\textbf{x}}{\ast}V=\dfrac{1}{\vert T\vert}\int \ldots\int_{\mathbb{T}^{n}}V \ d\textbf{x}
\] 
and which also corresponds to the \textit{a priori} uniform potential of the $n$-torus (or of the whole space $\mathbb{R}^{n}$) in the absence of charge source. We take for convenience $V_{m}=0$ as reference value of potentials.
In these conditions,
\begin{equation*}
V(\textbf{x})=-4\pi\sum_{\textbf{k}\in \lbrace0,1\rbrace^{n}}(-1)^{\vert\textbf{k}\vert}\Psi (\textbf{x}-a\textbf{k})
\end{equation*}
and after simplification using (\ref{FS}) and the electroneutrality of the torus, we obtain the exact expression of the potential of zero mean value as
\begin{equation}\label{pot1}
V(\textbf{x})=\dfrac{4\pi}{\vert T\vert}\sum_{\textbf{k}\in \lbrace0,1\rbrace^{n}}(-1)^{\vert\textbf{k}\vert}\int_{0}^{+\infty}\Theta (\mathbf{x}-a\textbf{k}\vert v) dv.
\end{equation}
As an application, in the case of the planar ($n=2$) NaCl crystal characterized by the charge distribution on the flat torus $\mathbb{T}^{2}=\mathbb{R}^{2}/2a\mathbb{Z}\times 2a\mathbb{Z}$,
\begin{equation}
\rho (x_{1},x_{2})=\delta(x_{1},x_{2})+\delta(x_{1}-a,x_{2}-a)-\delta(x_{1}-a,x_{2})-\delta(x_{1},x_{2}-a),
\end{equation}
i.e. a point charge $(-1)^{k+m}$ located at $(ka,mb)\in \mathbb{R}^{2}$, one finds after simplification and using a standard translation formula for the $\vartheta_{3}$ function, the following integral representation
\begin{gather}
V(x_{1},x_{2})=\int_{0}^{+\infty}\left\lbrace \vartheta_{3}\left( \left.\frac{\pi x_{1}}{2a} \right| iv\right)\vartheta_{3}\left( \left.\frac{\pi x_{2}}{2a} \right| iv\right)+\vartheta_{4}\left( \left.\frac{\pi x_{1}}{2a} \right| iv\right)\vartheta_{4}\left( \left.\frac{\pi x_{2}}{2a} \right| iv\right)
\right. 
\\ \notag \\ \notag
\left. -\vartheta_{4}\left( \left.\frac{\pi x_{1}}{2a} \right| iv\right)\vartheta_{3}\left( \left.\frac{\pi x_{2}}{2a} \right| iv\right)-\vartheta_{3}\left( \left.\frac{\pi x_{1}}{2a} \right| iv\right)\vartheta_{4}\left( \left.\frac{\pi x_{2}}{2a} \right| iv\right)\right\rbrace dv .
\end{gather}
Nevertheless, it is important to notice that thanks to (\ref{psi1}), this potential may have also the closed-form exact expression:
\begin{eqnarray}
V(x_{1},x_{2})&=&-4\pi \ \Psi \overset{x_{1},x_{2}}{\ast}\rho
\\ \notag
&=&\log \left| \dfrac{\vartheta_{1}\left( \left.\frac{\pi z}{2a} -\frac{\pi}{2}\right| i\right)\vartheta_{1}\left( \left.\frac{\pi z}{2a} -i\frac{\pi}{2}\right| i\right)}{\vartheta_{1}\left( \left.\frac{\pi z}{2a} \right| i\right)\vartheta_{1}\left( \left.\frac{\pi z}{2a} -\frac{\pi}{2}-i\frac{\pi}{2}\right| i\right)}\right|^{2}
\\ \notag
&=&\log \left| \dfrac{\vartheta_{2}\left( \left.\frac{\pi z}{2a} \right| i\right)\vartheta_{4}\left( \left.\frac{\pi z}{2a} \right| i\right)}{\vartheta_{1}\left( \left.\frac{\pi z}{2a} \right| i\right)\vartheta_{3}\left( \left.\frac{\pi z}{2a} \right| i\right)}\right|^{2} \ \ \ \ \mathrm{with} \ \ z=x_{1}+ix_{2}.
\end{eqnarray}

\subsection{CsCl crystal}

The $n$-dimensional CsCl-type crystal has a body-centered hypercubic structure characterized by the following charge distribution on the flat torus $\mathbb{T}^{n}$,
\[
\rho (\textbf{x})=\delta(\textbf{x})-\delta(\textbf{x}-a\textbf{1}), \ \ \ \textbf{1}=(1,1,\ldots 1)
\]
i.e. a cation Cs$^{+}$ at each lattice point and an anion Cl$^{-}$ at the center of the hypercubic unit cell. As done previously, one obtain from (\ref{AA}) and the fundamental solution (\ref{FS}), the following integral representation for the electrostatic potential of zero mean value,
\begin{equation}\label{BB}
V(\textbf{x})=\dfrac{4\pi}{\vert T\vert}\int_{0}^{+\infty}\lbrace \Theta (\mathbf{x}\vert v)-\Theta (\mathbf{x}-a\textbf{1}\vert v)\rbrace dv.
\end{equation}
In particular, for a square CsCl crystal structured as $\rho (x_{1},x_{2})=\delta(x_{1},x_{2})-\delta(x_{1}-a,x_{2}-a)$, both formulas are valid:
\begin{eqnarray}
V(x_{1},x_{2})&=&\int_{0}^{+\infty}\left\lbrace \vartheta_{3}\left( \left.\frac{\pi x_{1}}{2a} \right| iv\right)\vartheta_{3}\left( \left.\frac{\pi x_{2}}{2a} \right| iv\right)-\vartheta_{4}\left( \left.\frac{\pi x_{1}}{2a} \right| iv\right)\vartheta_{4}\left( \left.\frac{\pi x_{2}}{2a} \right| iv\right)\right\rbrace dv \notag
\\ \notag \\
&=&\log \left| \dfrac{\vartheta_{3}\left( \left.\frac{\pi z}{2a} \right| i\right)}{\vartheta_{1}\left( \left.\frac{\pi z}{2a} \right| i\right)}\right|^{2} \ \ \ \ \mathrm{with} \ \ z=x_{1}+ix_{2}.
\end{eqnarray}

\section{Madelung constant for $n\geq 3$}

\subsection{NaCl crystal}

Let us examine the potential (\ref{pot1}) at the origin $\mathbf{x}=\mathbf{0}$ (but the calculation should be similar at any other charge site $\mathbf{x}=a\mathbf{k_{0}}, \ \mathbf{k_{0}}\in \mathbb{Z}^{n}$) that we recast as,
\[
V(\textbf{0})=\dfrac{4\pi}{2^{n}a^{n}}\int_{0}^{+\infty}\left\lbrace \sum_{\textbf{k}\in \lbrace0,1\rbrace^{n}}(-1)^{\vert\textbf{k}\vert}\Theta (a\textbf{k}\vert v)\right\rbrace  dv .
\]
Taking in account the quasi-periodicity properties of theta functions and considering the definition (\ref{theta}), one may distinguish the following terms in the latter summation:
\begin{itemize}
\item $\vartheta_{3}^{n}(0\vert i\pi v/a^{2})$ for $\vert \textbf{k}\vert =0$;
\item $-n \ \vartheta_{3}^{n-1}(0\vert i\pi v/a^{2})\vartheta_{4}(0\vert i\pi v/a^{2})$ for $\vert \textbf{k}\vert =1$;
\item $\binom{n}{2}  \vartheta_{3}^{n-2}(0\vert i\pi v/a^{2})\vartheta_{4}^{2}(0\vert i\pi v/a^{2})$ for $\vert \textbf{k}\vert =2$;
\item \ldots
\item $(-1)^{n}\binom{n}{n} \vartheta_{4}^{n}(0\vert i\pi v/a^{2})$ for $\vert \textbf{k}\vert =n$.
\end{itemize}
Therefore, the latter summation is a binomial expansion and it is possible to write after a change of variable,
\begin{equation}\label{pot2}
V(\textbf{0})=\dfrac{1}{2^{n-2}a^{n-2}}\int_{0}^{+\infty} \left\lbrace\vartheta_{3}(0\vert i v)-\vartheta_{4}(0\vert i v) \right\rbrace^{n}dv. 
\end{equation}
Since (see e.g. \cite{Walker})
\begin{equation}\label{asympt}
\vartheta_{3}\left( 0\left| i v\right.\right)=\dfrac{1}{\sqrt{ v}}\left\lbrace 1+O(e^{-\pi /v})\right\rbrace \ \ \mathrm{and} \ \ \vartheta_{4}\left( 0\left| i v\right.\right)=\dfrac{2}{\sqrt{ v}}e^{-\pi /4v}\left\lbrace 1+O(e^{-2\pi /v})\right\rbrace, 
\end{equation}
the integrand is $O(v)^{-n/2}$ as $v$ tends to zero. It ensues that the integral (\ref{pot2}) is divergent for $n\geq 2$ and unsurprisingly  the potential is well infinite at the origin.

Let us consider now the \textit{Hadamard finite part} of the integral (\ref{pot2}) we define as follows for $n\geq 3$ (see e.g. \cite{Blanchet_Faye} for a comprehensive presentation),
\begin{eqnarray}\label{FP1}
V_{M}(\textbf{0})&:=&\dfrac{1}{2^{n-2}a^{n-2}}\mathrm{FP}\int_{0}^{+\infty} \left\lbrace\vartheta_{3}(0\vert i v)-\vartheta_{4}(0\vert i v) \right\rbrace^{n}dv
\\ \notag
&=&\dfrac{1}{2^{n-2}a^{n-2}} \ \underset{\epsilon \rightarrow 0+}{\lim}\int_{\epsilon}^{+\infty} \left[ \left\lbrace\vartheta_{3}(0\vert i v)-\vartheta_{4}(0\vert i v) \right\rbrace^{n}-\dfrac{1}{v^{n/2}}\right] dv 
\\ \notag
&=&\dfrac{1}{2^{n-2}a^{n-2}} \ \underset{\epsilon \rightarrow 0+}{\lim}\left( \int_{\epsilon}^{+\infty}  \left\lbrace\vartheta_{3}(0\vert i v)-\vartheta_{4}(0\vert i v) \right\rbrace^{n} dv -\dfrac{1}{\left( n/2 -1\right) \epsilon^{n/2-1}}\right) .
\end{eqnarray}
Remark then that the regularization term $(n/2-1)^{-1}\epsilon^{1-n/2}$ which removes the singularity from the integral may be interpreted as the $n$-Coulomb potential (zero at infinity) due to a solitary unit charge located at the origin, at a distance proportional to $\sqrt{\epsilon}$ (recall that such a potential is exactly written as $\Gamma (n/2)/(n/2-1)\pi^{n/2-1}\Vert \textbf{x}\Vert^{n-2}$). As a result, the finite part (\ref{FP1}) appears to be equal to the potential energy of the origin charge in the $n$-dimensional NaCl crystal i.e. the electrostatic energy binding this unit charge to the rest of the crystal which is usually always written as the infinite sum of contributions of charge sites,
\begin{equation}\label{lat5}
 V_{M}(\textbf{0})=\alpha (n)\sum_{\mathbf{k}\in \mathbb{Z}^{n}\setminus \mathbf{\lbrace 0\rbrace}}\dfrac{(-1)^{\vert\textbf{k}\vert}}{\Vert\mathbf{k}\Vert^{n-2}} \ \ \ \ \mathrm{with} \ \ \ \alpha (n)=\dfrac{\Gamma \left( n/2\right) }{(n/2-1)\pi^{n/2-1}a^{n-2}}.
 \end{equation} 
The finite part integral (\ref{FP1}) thus gives an exact formulation of the so-called Madelung constant $M_{\mathrm{NaCl}}=V_{M}(\textbf{0})$ for the NaCl crystal in any dimension $n\geq 3$ whose evaluation is far from raising the same fundamental mathematical difficulties than those unveiled by the conditionally convergent lattice sum (\ref{lat5}).

As an application for the $3$-dimensional NaCl crystal, if we adopt the common convention of the nearest-neighbour separation as unit length by setting $a=1$, one must have:
\[
M_{\mathrm{NaCl}}=\underset{\epsilon \rightarrow 0+}{\lim}\left( \dfrac{1}{2}\int_{\epsilon}^{+\infty}  \left\lbrace\vartheta_{3}(0\vert i v)-\vartheta_{4}(0\vert i v) \right\rbrace^{3} dv -\dfrac{1}{\sqrt{\epsilon}}\right)= \sum_{\vert k\vert +\vert m\vert +\vert p\vert \neq 0}\dfrac{(-1)^{k+m+p}}{\left( k^{2}+m^{2}+p^{2}\right)^{1/2} }.
\]
Choosing $\epsilon =0.000001$ and using a standard method of numerical integration in a Mathematica$^{\circledR}$ environment for instance, one easily finds for the finite part the value
\[
M=-1.747 \ 564 \ 594 \ 021 \ \ldots 
\]
which is exact up to the seventh decimal compared with (\ref{Madelung_NaCl}). We can hope to obtain more accurate results for smaller $\epsilon$ together with more refined numerical integration procedure.

It is worth to notice that the latter numerical value (or more generally the one we should calculate from (\ref{FP1})) is meaningful only in comparison to the potential zero reference: that of any point of the whole space void of charge sources or the value obtained at infinity for the $n$-Coulomb potential due to an isolated charge when $n\geq 3$. In addition, one can check that the electrostatic potential energy or the Madelung constant keeps the same value $M_{\mathrm{NaCl}}$ considering any charge site so that the total electrostatic energy per unit cell (i.e. a fundamental domain of the torus $\mathbb{T}^{n}$) is
\[
U=\dfrac{1}{2}\times 2^{n}M_{\mathrm{NaCl}}=2^{n-1}M_{\mathrm{NaCl}}.
\]
For $n=3$, $U\approx -6.990 \ 258 \ 4$.

\subsection{CsCl crystal}

In the same way from (\ref{BB}), one finds the divergent integral
\begin{equation}\label{pot3}
V(\textbf{0})=\dfrac{1}{2^{n-2}a^{n-2}}\int_{0}^{+\infty} \left\lbrace\vartheta_{3}(0\vert i v)^{n}-\vartheta_{4}(0\vert i v)^{n} \right\rbrace dv ,
\end{equation}
whose Hadamard finite part
\begin{equation}
M_{\mathrm{CsCl}}:=\dfrac{1}{2^{n-2}a^{n-2}} \ \underset{\epsilon \rightarrow 0+}{\lim}\left( \int_{\epsilon}^{+\infty}  \left\lbrace\vartheta_{3}(0\vert i v)^{n}-\vartheta_{4}(0\vert i v)^{n} \right\rbrace dv -\dfrac{1}{\left( n/2 -1\right) \epsilon^{n/2-1}}\right)
\end{equation}
defines the Madelung constant for the CsCl crystal in any dimension $n\geq 3$. Recall that this constant is usually computed by the lattice sum
\begin{equation}
M_{\mathrm{CsCl}}=\alpha (n)\left\lbrace \sum_{\mathbf{k}\in (2\mathbb{Z})^{n}\setminus \mathbf{\lbrace 0\rbrace}}\dfrac{1}{\Vert\mathbf{k}\Vert^{n-2}}-\sum_{\mathbf{k}\in (2\mathbb{Z}+1)^{n}}\dfrac{1}{\Vert\mathbf{k}\Vert^{n-2}}\right\rbrace ,
\end{equation}
where $\alpha(n)$ has been defined in (\ref{lat5}).

As an application for the $3$-dimensional CsCl crystal, if we choose the cubic unit cell side as unit length by setting $2a=1$ (for the nearest-neighbour convention, we should take $a=1/\sqrt{3}$), one finds using a standard numerical integration with $\epsilon=0.000001$,
\[
M_{\mathrm{CsCl}}=\underset{\epsilon \rightarrow 0+}{\lim}\left( \int_{\epsilon}^{+\infty}  \left\lbrace\vartheta_{3}(0\vert i v)^{3}-\vartheta_{4}(0\vert i v)^{3} \right\rbrace dv -\dfrac{2}{\sqrt{\epsilon}}\right)=-2.035 \ 361 \ 508 \ 229 \ \ldots \ .
\]
This result agrees up to the eighth decimal with the numerical values otherwise obtained for instance in \cite{Hautot,Zucker}. As indicated above, this constant is the electrostatic potential energy of each charge in the crystal so that the total electrostatic energy per unit cell (of side $1$) is for $n=3$,
\[
U=\dfrac{1}{2}\times 2M_{\mathrm{CsCl}}\approx -4.070 \ 723 \ 02.
\]

\section{Madelung constant for $n=2$}

Let us examine for ending the very singular case of the planar NaCl and CsCl crystal (i.e. in a 3-dimensional approach, an array of uniformly charged lines regularly spaced and perpendicular to the $x_{1},x_{2}$-plane, not to be confused with a simple lattice of point charges lying in the $x_{1},x_{2}$-plane of the 3-dimensional space ) which is, to our knowledge, never adressed in the literature. 
%
 
 \subsection{CsCl crystal} 
 
For clarity of our presentation, we begin in this section with the CsCl crystal characterized by the following charge distribution on the flat torus $\mathbb{T}^{2}=\mathbb{R}^{2}/2a\mathbb{Z}\times 2a\mathbb{Z}$,
\begin{equation}\label{distrib0}
\rho (x_{1},x_{2})=\delta (x_{1},x_{2})-\delta (x_{1}-a,x_{2}-a),
\end{equation}
i.e. point charges $+1$ located at each lattice point $2na+i2ma, \ (n,m)\in \mathbb{Z}^{2}$ and a negative point charge $-1$ at center of each unit cell. The potential at the origin is given by (\ref{pot3}) taking $n=2$:
 \[
V(0,0)= \int_{0}^{+\infty}  \left\lbrace\vartheta_{3}(0\vert i v)^{2}-\vartheta_{4}(0\vert i v)^{2} \right\rbrace dv .
 \]
 The Hadamard finite part of this divergent integral because of the logarithmic singularity must be here defined as
 \begin{eqnarray}\label{CC}
V_{M}(0,0)&:=&\mathrm{FP}\int_{0}^{+\infty} \left\lbrace\vartheta_{3}(0\vert i v)^{2}-\vartheta_{4}(0\vert i v)^{2} \right\rbrace dv
\\ \notag
&=&\underset{\epsilon \rightarrow 0+}{\lim}\left( \int_{\epsilon}^{A} \left[ \left\lbrace\vartheta_{3}(0\vert i v)^{2}-\vartheta_{4}(0\vert i v)^{2} \right\rbrace -\dfrac{1}{v}\right] dv +\int_{A}^{+\infty}  \left\lbrace\vartheta_{3}(0\vert i v)^{2}-\vartheta_{4}(0\vert i v)^{2} \right\rbrace dv\right) 
\\ \notag
&=&\underset{\epsilon \rightarrow 0+}{\lim}\left( \int_{\epsilon}^{+\infty}  \left\lbrace\vartheta_{3}(0\vert i v)^{2}-\vartheta_{4}(0\vert i v)^{2} \right\rbrace dv +\log \dfrac{\epsilon}{A}\right) 
\end{eqnarray}
where $A>0$ is an arbitrary parameter. Choosing appropriately the constant $A$, it is thus possible to assign any value to $V_{M}(0,0)$. Contrary to the case $n\geq 3$, there exists for $n=2$ an insuppressible ambiguity implying that the Madelung constant for a such planar crystal cannot be defined in this manner.

Now, let us notice that such an ambiguity (i.e. the arbitrariness in $A$) can also be related with the scale invariance of the Poisson equation $\Delta V=-4\pi q \delta(x_{1},x_{2})$ in the plane. Indeed, the scale-invariant solution of this equation is the potential due to the origin charge $q$ and expressed as
\begin{equation}\label{DD}
V(x_{1},x_{2})=-2q\log \dfrac{\vert z\vert}{R} \ \ \ , \ \ z=x_{1}+ix_{2} 
\end{equation}
where $R>0$ is an arbitrary parameter we have called \textit{horizon} in \cite{Mam} which allows ($i$) to adimensionalize the logarithm, ($ii$) to fix a (finite) distance between the charge and a reference point of the plane where the potential is zero. Compared with (\ref{CC}), we note thereby that the regularization term $\log \epsilon /A$ which removes the singularity from the integral is thus simply the self-potential of the unit origin charge at the distance $\sqrt{\epsilon}$ and of horizon $R=\sqrt{A}$ we can choose arbitrarily.

For defining the Madelung constant for the planar CsCl crystal, it is necessary to proceed differently. 

In contrast with the case $n\geq 3$ where the horizon of any charge considered separately is infinite, we begin to fix in the plane $\mathbb{R}^{2}$ the reference zero potential at the origin if there is no charge. Then, the potential at any point $z=x_{1}+ix_{2}$ of the plane due to the charge distribution (\ref{distrib0}), \textit{the origin charge being removed}, is thus given by the infinite double sums,
\begin{gather}
V_{M}(x_{1},x_{2})=-\sum_{(n,m)\in \mathbb{Z}^{2}\setminus \lbrace\textbf{0}\rbrace}\log\left\lbrace \dfrac{(x_{1}-2na)^{2}+(x_{2}-2ma)^{2}}{4n^{2}a^{2}+4m^{2}a^{2}}\right\rbrace \phantom{XXXXXX} \notag
\\
\phantom{XXXXXXX}+ \sum_{(n,m)\in \mathbb{Z}^{2}}\log\left\lbrace \dfrac{(x_{1}-(2n+1)a)^{2}+(x_{2}-(2m+1)a)^{2}}{(2n+1)^{2}a^{2}+(2m+1)^{2}a^{2}}\right\rbrace  . \label{pot6}
\end{gather}
Owing to the choice of the reference point, the latter formula obviously gives
\begin{equation}\label{zero}
V_{M}(\textbf{0})=V_{M}(0,0)=0
\end{equation}
i.e. the potential energy of the origin charge is zero. But the potential energy of any other charges of the unit cell is not zero as we shall see below so that the latter result \textit{must not be interpreted} as meaning that the Madelung constant for the crystal is null. In addition, this result also shows that the definition of the Madelung constant by a lattice sum like (\ref{lat5}) or relatives is problematic and illusive in the $2$-dimensional settings.

Nevertheless, following an approach outlined in \cite{Mam} for obtaining the fundamental solution for the Laplacian in a square, let us remark that the latter logarithmic series are each the logarithm of the square modulus of following functions of complex variable $z=x_{1}+ix_{2}$ expressed by the infinite double products,
\[
\phi_{1}(z)=\prod_{(n,m)\in \mathbb{Z}^{2}\setminus \lbrace\textbf{0}\rbrace}\left( 1+\dfrac{\pi z/2a}{n+im}\right) 
\]
and
\[
\phi_{2}(z)=\prod_{(n,m)\in \mathbb{Z}^{2}}\left( 1+\dfrac{\pi z/2a}{n-\frac{1}{2}+i(m-\frac{1}{2})}\right)
\]
respectively. Since these products are not absolutely convergent, a rearrangement of terms is necessary to make them convergent. Using the Eisenstein's summation convention (see \cite{Walker,Olver,NIST:DLMF,Mam}), a connection with Jacobi theta functions can be found such that
\[
\phi_{1}(z)=\dfrac{2a}{\pi z}\dfrac{\vartheta_{1}\left( \left.\frac{\pi z}{2a}\right| i\right)}{\vartheta'_{1}\left( 0 \vert i\right)} \ \ \ \ , \ \ \ \ \phi_{2}(z)=\dfrac{\vartheta_{3}\left( \left.\frac{\pi z}{2a}\right| i\right)}{\vartheta_{3}\left( 0 \vert i\right)} .
\]
Hence, (\ref{pot6}) has the closed-form expression,
\begin{equation}\label{pot8}
V_{M}(x_{1},x_{2})=\log \left|\dfrac{\vartheta_{3}\left( \left.\frac{\pi z}{2a} \right| i\right)}{\vartheta_{1}\left( \left.\frac{\pi z}{2a}\right| i\right)}\right|^{2}+2 \log \left( \dfrac{\pi}{2a}\dfrac{\vartheta'_{1}\left( 0 \vert i\right)}{\vartheta_{3}\left( 0 \vert i\right)}\vert z\vert\right) .
\end{equation}
Since the modular parameter of theta functions is equal to $i$ (the square lattice $\Lambda =2a\mathbb{Z}\times 2a\mathbb{Z}$ is a special important case in algebraic geometry and number theory known as a lemniscatic lattice), it is worth to note for further reference these stricking particular values (see e.g. \cite{Grad,Walker,Olver,NIST:DLMF}),
\begin{equation}\label{identity}
\vartheta_{3}(0\vert i)=\sqrt{\dfrac{2\textbf{K}}{\pi}}=2^{-1/2}\pi^{-3/4}\Gamma (1/4) \ \ \ , \ \ \ \vartheta_{2}(0\vert i)=\vartheta_{4}(0\vert i)=2^{-1/4}\vartheta_{3}(0\vert i)
\end{equation}
together with the Jacobi's identity:
\[
\vartheta_{1}'(0\vert i)=\vartheta_{2}(0\vert i)\vartheta_{3}(0\vert i)\vartheta_{4}(0\vert i)=2^{-1/2}\vartheta_{3}(0\vert i)^{3} .
\]
$\mathbf{K}$ denotes here as usual the complete elliptic integral $\mathbf{K}=\mathbf{K}(1/\sqrt{2})=\pi \vartheta_{3}(0\vert i)^{2}/2$.

As prescribed by the definition (\ref{pot6}), the potential (\ref{pot8}) is well equal at any point $(x_{1},x_{2})$ of the plane to:
\begin{itemize}
\item the electrostatic potential due to the whole crystal which could be readily calculated from (\ref{psi1}) and (\ref{distrib0}) as the double convolution product on the flat torus $\mathbb{R}^{2}/ 2a\mathbb{Z}\times 2a\mathbb{Z}$ (recall that a unit cell of the crystal thus corresponds to a square domain of the torus),
\begin{equation}\label{logar2}
V(x_{1},x_{2})=-4\pi \ \Psi \overset{x_{1},x_{2}}{\ast}\rho =\log \left|\dfrac{\vartheta_{3}\left( \left.\frac{\pi z}{2a} \right| i\right)}{\vartheta_{1}\left( \left.\frac{\pi z}{2a}\right| i\right)}\right|^{2}
\end{equation}
\item from which the self-potential of the origin charge
\[
-2 \log \left( \dfrac{\mathbf{K}}{a\sqrt{2}}\vert z\vert\right) 
\]
is removed, its horizon being thus fixed once and for all to the value $R=a\sqrt{2}/\mathbf{K}$.
\end{itemize}
 
Since
\begin{equation}\label{logar1}
V(x_{1},x_{2})= -2\log\left( \dfrac{\mathbf{K}}{a\sqrt{2}}\vert z\vert\right) +O(\vert z\vert)  \ \ \ \ \ \ \mathrm{as} \ \ \ \ \ \ \vert z\vert \rightarrow 0+,
\end{equation}
one finds obviously $\underset{ z \rightarrow 0}{\lim}V_{M}(x_{1},x_{2})=V_{M}(0,0)=0$
which fully confirms that the origin has been taken as the reference zero potential and accordingly, the potential energy of the origin charge is zero.

But, owing to the periodicity of the crystal, this choice of the potential reference point implies that \textit{the potential which binds each cation Cs$^{+}$ located at a lattice point, to the rest of the crystal is also zero} while for the anion Cl$^{-}$ at the center of the unit cell this potential may be evaluated as
\[
V_{M}(a,a)=\underset{ z \rightarrow a+ia}{\lim}\left[ V(x_{1},x_{2})-2\log  \dfrac{\vert z-a-ia\vert}{a\sqrt{2}}\right] =2\log \mathbf{K}
\]
i.e. by removing from (\ref{logar2}) the self-potential of the anion with the horizon $a\sqrt{2}$.

As a result, the total electrostatic energy per unit cell is exactly
\[
U=\dfrac{1}{2}\sum q V_{M}=-\log \mathbf{K} =-\log \left( \dfrac{\Gamma (1/4)^{2}}{4\sqrt{\pi}}\right) .
\]
This leads us to define the Madelung constant related to the 2D CsCl crystal as the \textit{mean value of potential energies of the two charges} which compose the unit cell viz.
\[
M_{\mathrm{CsCl}}=U=-\log \mathbf{K}=-0.617 \ 385 \ 745 \ 351 \ 564 \ldots \ .
\]
It is worth to notice that all previous quantities do not depend on the lattice parameter $a$ i.e. on the actual size of the square unit cell in accordance with the scale invariance for this $2$-dimensional electrostatic problem.
 
 \subsection{NaCl crystal} 
 
Consider the charge distribution on the flat torus $\mathbb{T}^{2}=\mathbb{R}^{2}/2a\mathbb{Z}\times 2a\mathbb{Z}$,
\begin{equation}\label{distrib}
\rho (x_{1},x_{2})=\delta (x_{1},x_{2})+\delta (x_{1}-a,x_{2}-a)-\delta (x_{1}-a,x_{2})-\delta (x_{1},x_{2}-a),
\end{equation}
i.e. point charges $(-1)^{n+m}$ located at $na+ima, \ (n,m)\in \mathbb{Z}^{2}$. As above, the potential reference point is taken at the origin for all charges except for the origin charge $+1$. Recall that this means the potential energy of this origin charge is zero and owing to the periodicity of the crystal structure in the variables $x_{1}, \ x_{2}$ and along the diagonal of the unit cell, this implies also that the potential energy of all cations Na$^{+}$ is zero. 

Since the potential within the crystal may be obtained in closed-form from (\ref{psi1}) and (\ref{distrib}) as the convolution product
\begin{eqnarray}\label{logar3}
V(x_{1},x_{2})&=&-4\pi \ \Psi \overset{x_{1},x_{2}}{\ast}\rho
\\ \notag
&=&\log \left|\dfrac{\vartheta_{1}\left( \left.\frac{\pi z}{2a}-\frac{\pi}{2} \right| i\right)\vartheta_{1}\left( \left.\frac{\pi z}{2a}-i\frac{\pi}{2} \right| i\right)}{\vartheta_{1}\left( \left.\frac{\pi z}{2a}\right| i\right)\vartheta_{1}\left( \left.\frac{\pi z}{2a}-\frac{\pi}{2}-i\frac{\pi}{2} \right| i\right)}\right|^{2}
\\ \notag \\ \notag
&=&\log \left|\dfrac{\vartheta_{2}\left( \left.\frac{\pi z}{2a} \right| i\right)\vartheta_{4}\left( \left.\frac{\pi z}{2a} \right| i\right)}{\vartheta_{1}\left( \left.\frac{\pi z}{2a} \right| i\right)\vartheta_{3}\left( \left.\frac{\pi z}{2a} \right| i\right)}\right|^{2} \ \ \ \ \ \ \mathrm{with} \ \ \ \ \ \ z=x_{1}+i x_{2} ,
\end{eqnarray}
it is no more difficult to derive the potential which binds each anion Cl$^{-}$ of the unit cell (and of horizon $a$) to the crystal as the limit (see (\ref{identity}) for a simplification)
\[
V_{M}(0,a)=V_{M}(a,0)=\underset{ z \rightarrow a}{\lim}\left[ V(x_{1},x_{2})-2\log  \dfrac{\left| z-a\right|}{a}\right] =2\log \mathbf{K} .
\]
Hence, the total electrostatic energy per unit cell is
\[
U=\dfrac{1}{2}\sum q V_{M}=-2\log\mathbf{K}.
\]
Again the Madelung constant is obtained as mean value of potential energies of the four charges which belong to the unit cell and we find
\[
M_{\mathrm{NaCl}}=\dfrac{U}{2}=-\log\mathbf{K}=M_{\mathrm{CsCl}}=-0.617 \ 385 \ 745 \ 351 \ 564 \ldots \ .
\]
One may ask why a such equality is obtained. The reason is simply that both planar crystals are structurally identical: indeed, up to a scale factor, one is obtained from the other by rotating $45^{\circ}$ around the origin. Due to the scale invariance, the Madelung constant is a physical invariant characteristic of the crystal structure, and is well identical in both cases.

\section{Conclusions}

In contrast with its common definition by lattice sums conditionally and slowly convergent which are often unsuitable for numerical computation, we have shown that the Madelung constant for $n\geq 3$-dimensional cubic crystal lattices may be obtained in an efficient way by means of (the finite part of) an integral of modular functions involving Jacobi theta functions. This definition is general and provides in Gaussian units an unique, exact and unambiguous presentation of this physical constant for any type of hypercubic crystal structured by point charges. The calculation for  NaCl and CsCl crystals has shown that such an approach is fully coherent with traditional ones and gives highly accurate results by the implementation of a standard numerical integration scheme.

The $n=2$ case (lemniscatic lattice) is apart in two respects due to the scale invariance of logarithmic Coulomb potentials: ($i$) such a property requires to fix, at finite distance, a potential reference point (chosen as the origin of the Euclidean plane) and thus leads to the impossibility to define the Madelung constant both by lattice sums and the regularization method as proposed above; ($ii$) contrary to crystals of higher dimension, a closed-form exact expression of the doubly periodic potential within the NaCl/CsCl crystal is nonetheless obtained and has allowed to define the Madelung constant as the mean value of potential energies of point charges composing the unit cell. In agreement with the scale invariance, this physical constant does not depend on the actual size of the crystal but solely on its invariant geometric structure.

\section*{Acknowledgements}

The author would like to thank the referees for their valuable remarks and suggestions, and his colleague Thierry Mara for many stimulating discussions.

\bibliographystyle{unsrt}
\bibliography{biblio1}

\end{document}